\documentclass[
    ,final            
  ]
  {aipproc}

\layoutstyle{6x9}

\begin{document}

\title{Heavy Ion Collisions and New Forms of Matter}

\classification{21.65.Qr, 12.38.-t}
\keywords      {High Energy Density Matter, Heavy Ion Collisions}

\author{Larry McLerran}{
  address={Riken Brookhaven Center and Physics Department, Brookhaven National
  Laboratory, Upton, NY 11973 USA}
}

\begin{abstract}
 I discuss forms of high energy density matter in QCD.  These include the Color Glass Condensate,
 the Glasma and the Quark Gluon Plasma.  These all might be studied in ultra-relativistic heavy ion collisions, and the Color Glass Condensate might also be probed in electron-hadron collisions.
I present the properties of such matter, and some aspects of what is known of their properties.
 \end{abstract}

\maketitle

\section{Introduction}

\begin{figure}[ht]
        \includegraphics[width=0.90\textwidth]{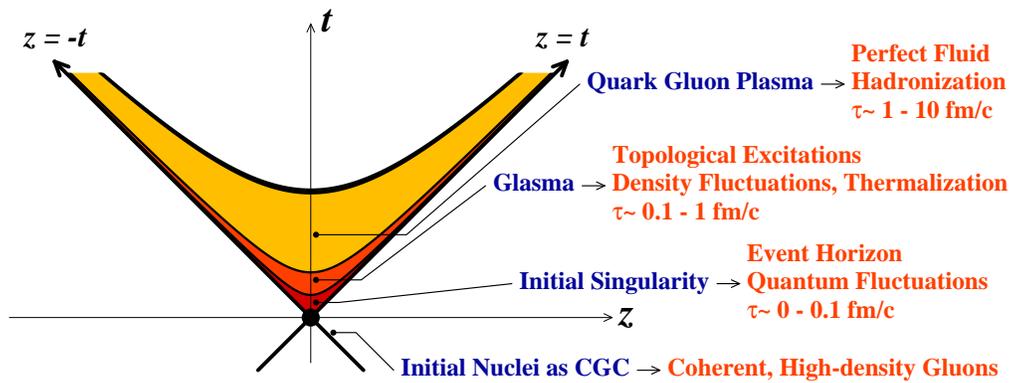}
        \caption{A schematic picture of the evolution of matter produced in the heavy ion collisions. }
\label{lilbang}
\end{figure}

There is a generic similarity between the physical conditions in the early stages of heavy ion collisions, and those of the early universe.  This is illustrated in Fig. \ref{lilbang} . The Color Glass Condensate
describes the initial quantum mechanical wavefunctions which initiate the collision.  This is similar to the initial wavefunction 
in cosmology.
At the moment of the collision there is the initial singularity, analogous to that of the quantum gravity  and inflationary stages of cosmology, where quantum fluctuations are important. 
These fluctuations  eventually evolve into the large scale fluctuations in the matter
distribution produced in such collisions, analogous to the fluctuations produced in inflationary cosmology which form the seeds for matter fluctuations which eventually develop galaxies and clusters of galaxies. After the singularity there
is a Glasma phase which has topological excitations analogous to those associated with
baryon number violation in electroweak theory. The Quark Gluon Plasma is analogous to both the
electroweak and QCD thermal phases of expansion in cosmology, and the deconfinement transition
is analogous to that which generates masses for electroweak bosons.
In cosmology there are a variety of phase transitions which occur at various times, corresponding to the confinement-deconfinement transition
of QCD.     

There is a new paradigm:  Various forms of matter control
the high energy limit of QCD.  These forms of matter have intrinsically interesting properties, as well
as properties which once understood have extensions to other areas of physics such as cosmology.

It is the purpose of this talk to describe for you the properties of these high energy density forms of matter, how they appear in high energy physics phenomenology, and what is known experimentally about these forms of matter. For reasons of space, these topics are covered only in the most generic terms.

\section{The  Initial Wavefunction}

\begin{figure}[ht]
        \includegraphics[width=0.90\textwidth]{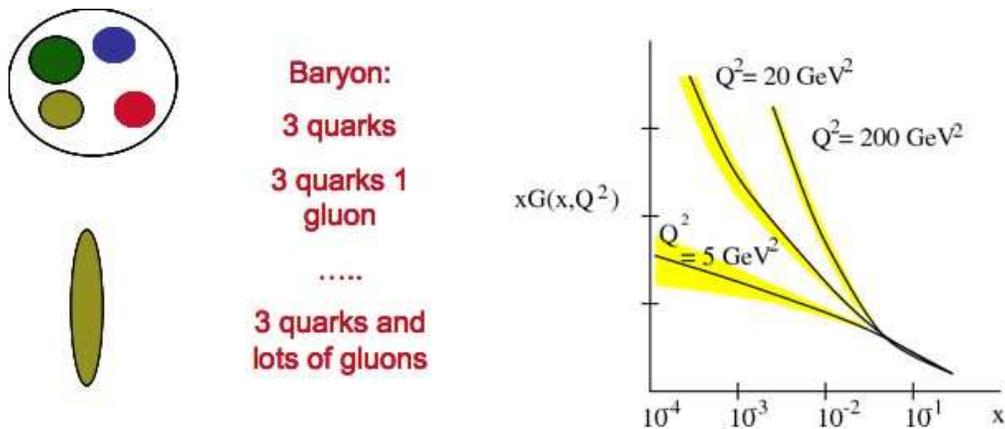}
        \caption{The leftmost figure on this slide illustrates the wavefunction for a gluon as a function
        of energy.  The rightmost figure illustrates the experimentally measured gluon distribution
        functions as a function of x. }
\label{wfn_glue}
\end{figure}

In the leftmost part of Fig. \ref{wfn_glue},  I show the  various Fock space components for
a baryon wavefunction.  At low energies, the dominant states for physical processes have three quarks and a few gluons.  In high energy collisions, many particles are produced.  These ultimately arise from 
components of the hadron wavefunction which have many gluons in them.  These components make
a gluon wall of longitudinal extent of $1/\Lambda_{QCD}$.  The density of gluons in the
transverse plane grows as the energy increases, as is also shown Fig. \ref{wfn_glue}

These gluons must become very tightly packed
together.  They form a high density, highly coherent condensate of gluons, the Color Glass Condensate (CGC).\cite{cgc1}-\cite{cgc3}
Because the typical separation of gluons is small, at high enough energy $\alpha_S <<1$,
and the system is weakly coupled. Due to the coherence, it is also strongly interacting.
An example of coherence greatly amplifying a very weak interaction is given by gravity:  The intrinsic
strength of the gravitational force is very small, but due to the coherent superposition of 
forces arising from individual particles, it becomes a large force.
 
There is a typical momentum scale $Q_{sat}$, and gluons with momentum less
than this scale have maximal phase space density,
\begin{equation}
	{{dN} \over {d^2p_Td^2r_Tdy}} \sim {1 \over \alpha_S}
\end{equation}
As one goes to higher energies, the saturation momentum increases, as more gluons are added to
the CGC.

The theory of the CGC has provided both a rich phenomenology as well as first principles understanding within QCD for the high energy limit. \cite{cgc3} Some of the successes of this theory include
\begin{itemize}
\item{Scaling properties in electron-hadron scattering.}
\item{Elastic and almost elastic electron-hadron scattering.}
\item{Nuclear size dependence of quark and gluon distributions.}
\item{Distributions of produced particles in hadron and nuclear collisions.}
\item{Scaling properties of hadron-hadron collisions.}
\item{Long range momentum correlations for particles produced in hadron-hadron collisions.}
\item {Intuitive understanding of high energy limit for cross sections.}
\item{The origin of the multiparticle excitations which control high energy scattering (Pomeron, Reggeon,
Odderon).}
\end{itemize}
This is a very active area of research.  It has as one of its intellectual
goals, the unification of the description of all strong interaction processes at high energy
and as such involves different phenomena:  electron-hadron, hadron-hadron, hadron-nucleus and nucleus-nucleus collisions. 

\section{The Initial Singularity and the Glasma}

Before the collision of two hadrons, two sheets of Colored Glass approach one another.  Because
the phase space density of gluons is large, the gluons can be treated as classical fields.  The fields
are Lorentz boosted Coulomb fields, that is Lienard-Wiechart potentials, which are static in the
transverse plane of the hadrons and have $E \perp B \perp \hat{z}$, where $z$ is the direction of motion.\begin{figure}[ht]
        \includegraphics[width=0.50\textwidth]{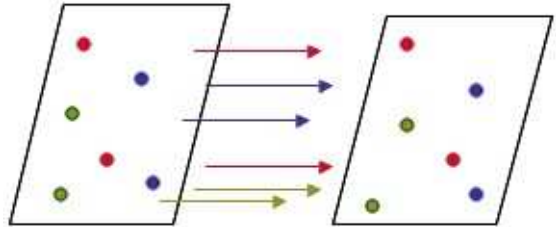}
        \caption{After the collision of two sheets of colored glass, longitudinal electric and
        magnetic fields associated with the Glasma form. }
\label{glasma}
\end{figure}
When the classical equations for the evolution of these fields is solved, in a very short 
time scale of order
$\Delta t \sim {1 \over Q_s} e^{-\kappa /\alpha_S}$, the fields change from purely transverse to purely
longitudinal.  This is because the collisions generate color electric and magnetic monopole charge
densities of opposite sign on the two sheets.  This description is reminiscent of the  flux tube models of
color electric fields which are used in phenomenological descriptions of low energy strong interactions.
At high energies,  there are both color electric and magnetic field because the electric and magnetic
fields were of equal magnitude in the CGC, and because of the electric-magnetic duality of QCD.

Fields with a non-zero $\vec{E} \cdot \vec{B}$ carry a topological charge.  In QCD, they are associated
with anomalous mass generation and chiral symmetry violation.  In electroweak theory, such
fields may be responsible for generating the baryon asymmetry of the universe.  In QCD, they may
generate the masses of those particles which constitute the visible matter of the universe. Each field
configuration violates CP.  An experimental discovery of the effects of such fields would be of great importance.
Theoretical ideas for experimental signatures are sketchy.\cite{tytgat}   These initial longitudinal fields
at the time of  production evolve into transverse fields and eventually form a Quark Gluon Plasma. The matter from production until the formation of the Quark Gluon Plasma is called the Glasma.\cite{glasma1}-\cite{glasma3}

The initial Glasma fields are unstable, and after a time scale of order $1/Q_s$,
instabilities begin to become of the order of the original classical fields. \cite{mrowcz}-\cite{romatschke}The origin of these Weibel instabilities was originally found
for plasmas close to thermal equilibrium by Mrowczinski.  Quantum fluctuations in the original
wavefunction can grow by these instabilities, and eventually overwhelm the longitudinal electric
and magnetic Glasma fields.  Perhaps these fields form a chaotic or turbulent liquid which might
thermalize and isotropize the system.\cite{mueller}

The amplification of quantum fluctuations to macroscopic magnitude is reminiscent of inflation
in the early universe.   These quantum fluctuations expand to size scale larger than the event
horizon during inflation, and are imprinted into the fabric of space-time.  At much later times, the
event horizon size scale becomes of the order of galactic size scales.  Ultimately,  these fluctuations drive gravitationally unstable modes which form galaxies and clusters of galaxies.

In heavy ion collisions, analogous fluctuations in the hadronic wavefunction might
appear as momentum dependent fluctuations.  Such
fluctuations might be frozen into the final state distribution of particles.  

\section{The Strongly Interacting Quark Gluon Plasma}

Some of the first data from RHIC showed that there was strong radial and elliptic flow, and strong jet
quenching. \cite{whitepaper} The flow data could be described using perfect fluid hydrodynamic equations.  Although
there is a good deal of uncertainty due to lack of precise knowledge of the initial state,
final decoupling, equations of state and the magnitude of viscous corrections, it is nevertheless
remarkable that  perfect fluid hydrodynamic computations can describe the data. 
Such a description was not possible for data at lower energies. \cite{whitetheory}
Such agreement involves comparison of almost limitless numbers of
transverse and longitudinal momentum distributions of produced particles as a function of the impact parameter of the
nuclear collisions.  The impact parameter of such collisions is determined by measurements of
the particle multiplicity both in the forward and central regions of the collision (at low longitudinal and high longitudinal momentum in the center of mass frame).
The agreement between data and experiment
seems to be improved by the more careful treatments of late time evolution in the form of hadrons,
and using Color Glass initial conditions.\cite{whitetheory}  

Perfect fluid hydrodynamics is the limit where interactions are very small and viscosities are zero.
This suggests that the Quark Gluon Plasma is well thermalized and fairly strongly interacting.  The degree to which the viscosity may be treated as zero is subject to a better understanding of the
initial conditions.  Color Glass Condensate initial condition allow for larger viscosities than the standard
phenomenological choices.    

Jet quenching is the disappearance of high transverse momentum particles because of interactions with the produced media.   The production of high transverse momentum particles is described well within perturbative QCD.  One needs only know the distribution functions of quarks and gluons inside hadrons and nuclei.  This can be done by deep inelastic scattering combined with additional data on hadron-nucleus collisions..  

There is  suppression of jet production in gold-gold collisions by a factor of four out to transverse momenta of about $20~GeV$, corresponding to several $GeV$.  Such a large energy loss is stunning and was not anticipated.\cite{whitepaper}, \cite{jets} 

These observations have led to a consensus that one has produced a strongly interacting Quark
Gluon Plasma, and that to a fair or good or perhaps excellent approximation, the system is thermalized.
A well thermalized system is a perfect fluid.  The outstanding issue is how perfect is the perfect fluid.

This may be parameterized by the dimensionless ratio of viscosity to entropy density, $\eta/s$.
As argued by Son and collaborators,\cite{son} $N=4$ supersymmetric
Yang-Mills theory  satisfies a bound
\begin{equation}
	{\eta \over s} \ge {1 \over {4\pi}}
\end{equation}	
with the lower bound approached for infinite coupling.
This is a beautiful result which leads to new insight.  Although $N=4$ supersymmetric theories bear
no direct relation to our physical world, this bound was also more generally argued as 
a consequence that in the strong coupling limit of QCD,  the limiting small
viscosity is reached when  the mean free path of a particle become equal to its deBroglie wavelength.  
Below
this limit, scattering is difficult due to quantum mechanical coherence.

Is $\eta/s$ close to the minimum value for temperatures accessible at RHIC in the QGP?  
There are reasons to be skeptical: (1)  Some computations  argue that one does not need extremely large cross sections within transport computations to describe the elliptic flow data.\cite{ko}   (2) Flow results depend on initial conditions, about which there is not yet
consensus, the way that the equation of state is treated, and how one coalesces quarks and gluons into
hadrons.  There is much work now on initial conditions,
so this has potential for resolution.  If the initial conditions are more like those predicted from the
Color Glass Condensate, one generates more elliptic flow, which could be reduced by increasing the viscosity. \cite{dumitru}(3)
New realistic computations with
relativistic hydro and viscosity are being done.  The preliminary
results from these hydro computations indicate that varying viscosity can be compensated
by various other uncertainties in the computations.\cite{heinz}-\cite{romatschke1}

\section{Strings and AdSCFT}

There has been some discussion at the literature about string theory and its relationship to heavy ion 
physics.\cite{son},\cite{others}  AdSCFT is a mathematical trick which allows one to compute the properties of $N=4$ supersymmetric Yang-Mills theory in terms of gravitational theory in curved space.  
The strong coupling limit of N=4 SUSY corresponds to weak coupling gravity, allowing strong
coupling computations for the Yang-Mills theory.
 
N= 4 supersymmetric Yang-Mills is not QCD.  There are many caveats:
\begin{itemize}
\item{ It has no mass scale and is conformally invariant.}
\item{It has no confinement and no running coupling constant.}
\item{It is supersymmetric.}
\item{It has no chiral symmetry breaking or mass generation.}
\item{It has six scalar and fermions in the adjoint representation.}
\end{itemize}
The interesting applications of this correspondence for QCD are in the strong
coupling limit. 
Even in lowest order strong coupling computations it is very speculative to make relationships
between this theory and QCD, because of the above.  It is much more difficult to relate
non-leading computations to QCD.  

It may be possible to correct some or all of the above problems, or, for various physical problems,
some of the objections may not be relevant.  As yet there is neither consensus nor compelling argument
for the conjectured fixes or phenomena which would insure that the $N=4$ supersymmetric Yang Mills results
would reliably reflect  QCD.

Further, these computations are applied to the QGP at temperatures above the deconfinement 
transition temperature, where the validity of
strong coupling limit is arguable.  Lattice computations give reliable computations of the 
properties of the QGP, and indicate the coupling is of intermediate to weak strength in this region.
There are improved weak coupling computations which agree
with lattice data at temperatures more than a few times $T_c$.\cite{blaizot}

The AdSCFT correspondence, is probably best thought of as a discovery tool with limited
resolving power.  An example is the $\eta/s$ computation.  The discovery of the bound on $\eta/s$
could be argued be verified by an independent argument, as
a consequence of  the deBroglie wavelength of particles becoming of the order of mean free paths.  
It is a theoretical discovery but its direct applicability to heavy ion collisions remains to be shown.

Certainly,  computations in lattice gauge theory
or in weak coupling provide reliable QCD computations of a limited set of
physical phenomena.  One has  had  hope
for analytic methods in  the strong coupling dynamics of QCD, and we should continue to have such hope.


\begin{theacknowledgments}
I gratefully acknowledge my colleagues Edmond Iancu, Dima Kharzeev, Genya Levin, Al Mueller, and Raju Venugopalan for their insights in formulating the material presented here.  

This manuscript has been authorized under Contract No. DE-AC02-98CH0886 with the US Department of Energy.

\end{theacknowledgments}


\begin{thebibliography}{9}

\bibitem{cgc1}  L. D. McLerran and R. Venugopalan, {\it
Phys. Rev. } {\bf D49}, 2233(1994); 3352 (1994); {\bf D50}, 2225 (1994).

\bibitem{cgc2} E. Iancu, A. Leonidov
and L. D. McLerran, {\it Nucl. Phys.} {\bf A692}, 583 (2001);
E. Ferreiro E. Iancu, A. Leonidov
and L. D. McLerran, {\it Nucl. Phys.} {\bf A710},373 (2002);J. Jalilian-Marian, 
{\it Phys. Rev,.} {\bf C70} 027902 (2004) E. Iancu and L. McLerran, {\it  Phys. Lett.}
{\bf B510}, 145 (2001);
J.~Jalilian-Marian, A.~Kovner, L.~McLerran and H.~Weigert,
{\it Phys.\ Rev.}\ {\bf D55} (1997), 5414; J.~Jalilian-Marian, A.~Kovner, A.~Leonidov and  H.~Weigert,
{\it Nucl.\ Phys.}\ {\bf B504} (1997), 415;
{\it Phys.\ Rev.}\ {\bf D59} (1999), 014014.


\bibitem{cgc3}
E. Iancu and R. Venugopalan, Publisehd in QGP3, Eds. R. C. Hwa and X. N. Wang, World Scientific.  hep-ph/0303204

\bibitem{tytgat} D. Kharzeev, R. D. Pisarski and M. H. G. Tytgat, {\it Phys. Rev. Lett.} {\bf 81}, 512 (1998);
hep-ph/9804221

\bibitem{glasma1}A. Kovner, L. McLerran and H. Weigert, {\it Phys. Rev. } {\bf D52} 3809 (1995);
6231 (1995). 


\bibitem{glasma2}A. Krasnitz and R. Venugopalan,
{\it Nucl. Phys. } {\bf B557} 237 (1999); {\it Phys. Rev. Lett. } {\bf 84} 4309 (2000); {Phys. Rev. Lett. } {\bf 86} 1717 (2001); A. Krasnitz, Y. Nara and R. Venugopalan,
{\it Phys. Rev. Lett.} {\bf 87} 192302 (2001); {\it Nucl. Phys. } {\bf A717} 268 (2003).



\bibitem{glasma3} T. Lappi, {\it Phys. Rev. } {\bf C67} 054903 (2003); T. Lappi and L. Mclerran, {\it Nucl. Phys. } {\bf A772} 200 (2006).


\bibitem{mrowcz} S. Mroczynski, {\it Phys. Lett.} {\bf B214} 587 (1988); {\bf B314} 118 (1993);
{\bf  B363}, 26 (1997).

\bibitem{strikland} P. Romatschke and M. Strikland, { \it Phys. Rev. } {\bf D68} 036004 (2003);
{\bf D70} 116006 (2004).


\bibitem{arnold} P. Arnold, J. Lenaghan, and G. Moore,  {\it JHEP} {\bf 0308} 002 (2003); 
 P. Arnold, J. Lenaghan, G. Moore and L. Yaffe, {\it Phys. Rev. Lett. } {\bf 94} 
072302 (2005).  



\bibitem{romatschke} P. Romatshke and R. Venugopalan, {\it Phys. Rev. Lett.} {\bf 96} 062302 (2006) .
 A. Dumitru and Y. Nara,  {\it JHEP} {\bf 0509} 041 (2005);
A. Dumitru, Y. Nara and M. Strikland,  {\it Phys. Lett. } {\bf B621} 89 (2005).


\bibitem{mueller} M. Asakawa, S. A. Bass and B. Muller, {\it Porg. Theor. Phys.} {\bf 116}, 725 (2007);
{\it Phys. Rev. Lett.} {\bf 96}, 252301 (2006).

\bibitem{whitepaper}

Reports of the Star, Phenix, Phobos and Brahms experimental collaborations
 in {\it Nucl. Phys} {\bf A757} (2005)
 Brahms Collaboration p  1;  Phobos Collaboration p 28; Star Collaboration p 102; Phenix Collaboration 184.


\bibitem{whitetheory}
For a theoretical review of the properties of the QGP and it effects in heavy ion
collisions see Workshop on New Discoveries at RHIC: The Current Case for
the Strongly Interactive  Quark Gluon Plasma, Brookhaven, May 14-15 , 2004,  {\it Nucl. Phys.} {\bf A750} (2005) .


\bibitem{jets} J. D. Bjorken, Fermilab-Pub-82/59-THY (1982) and erratum.  (unpublished);
D. Appel, {\it Phys. Rev. } {\bf D33} 717 (1986).

\bibitem{son} G. Policastro, D. T. Son and A. O. Starinets, {\it Phys. Rev. Lett.} {\bf 87} 081601 (2001);
P. Kovtun, D. T. Son and A. O. Starinets, {\it Phys. Rev. lett.} {\bf 94} 11601 (2005).

\bibitem{ko} V. Greco, C. M. Ko and P. Levai, {\it Phys. Rev. } {\bf C68} 034904 (2003).

\bibitem{dumitru}
H. Drescher, A. Dumitru, C. Gombeaud and J. Ollitrault, arXiv:0704.3353[nucl-th].

\bibitem{heinz} U. Heinz, private communication.

\bibitem{romatschke1} P. Romatschke and U. Romatschke, arXiv:0706.1522[nucl-th].

\bibitem{others} I. R. Klebanov, {\it Nucl. Phys.} {\bf B496} 231 (1997).

\bibitem{blaizot} J. P. Blaizot and E. Iancu, {\it Phys. Rev. Lett.} {\bf 70}, 3376 (1993);
{\it Nucl. Phys.} {\bf B417}, 608 (1994); J. P. Blaizot, E. Iancu and A. Rebhan, {\it Phys. Rev. lett.} {\bf 83}, 2906 (1999); {\it Phys. Lett.} {\bf B470}, 181 (1999); {\it Phys. Rev.} {\bf D63}, 065003 (2001).


\end{thebibliography}
\end{document}